\documentclass[aps,superscriptaddress,twocolumn,nofootinbib]{revtex4}



\usepackage{natbib}
\usepackage{graphicx}
\usepackage{dcolumn}
\begin{document}
% You should use BibTeX and apsrev.bst for references
%\bibliographystyle{apsrev}

%\preprint{}

\title{Market reaction to temporary liquidity crises and the permanent market impact}
% Optional argument for running titles on pages
%\title[]{}



\author{Adam Ponzi}
\email{ponzi@lagash.dft.unipa.it}
\affiliation{INFM-CNR, Unit\`a di Palermo, Palermo, Italy}

\author{Fabrizio Lillo}
\email{lillo@lagash.dft.unipa.it}
\affiliation{INFM-CNR, Unit\`a di Palermo, Palermo, Italy}
\affiliation{Dipartimento di Fisica e Tecnologie Relative, University of Palermo, Viale delle Scienze, Edificio 18, I-90128 Palermo, Italy}
\affiliation{Santa Fe Institute, 1399 Hyde Park Road, Santa Fe, NM 87501}

\author{Rosario N. Mantegna}
\email{mantegna@unipa.it}
\affiliation{INFM-CNR, Unit\`a di Palermo, Palermo, Italy}
\affiliation{Dipartimento di Fisica e Tecnologie Relative, University of Palermo, Viale delle Scienze, Edificio 18, I-90128 Palermo, Italy}

\date{\today}

\begin{abstract}
We study the relaxation dynamics of the bid-ask spread and of the
midprice after a sudden, large variation of the spread, corresponding
to a temporary crisis of liquidity in a double auction financial
market. We find that the spread decays very slowly to its normal value
as a consequence of the strategic limit order placement of liquidity
providers. We consider several quantities, such as order placement
rates and distribution, that affect the decay of the spread. We
measure the permanent impact both of a generic event altering the
spread and of a single transaction and we find an approximately linear
relation between immediate and permanent impact in both cases.
\end{abstract}

%\pacs{}
%\keywords{}

\maketitle
%\tableofcontents

\section{Introduction}

Many theoretical and empirical studies have examined microstructure
properties of double auction financial markets. Limit order book data
provide the maximum amount of information at the lowest aggregation
level. Early examples of investigations of order book data can be
found in \cite{Harris96,Hamao95,Biais95}. Other examples of
investigations using limit order books are the developing of
econometric models of limit order execution times \cite{Lo02} and the
empirical investigation of order aggressiveness and trader' order
submission strategy in an open limit order book \cite{Hall06}. These
microstructure studies are important both for analysing and modeling
the dynamics of the limit order book and in the investigation of
determinants of key financial variables such as bid-ask spread
\cite{Huang97}. Moreover, the statistical regularities observed in
these investigations can provide empirical and modeling support or
falsify conjectures about the origin of stylized facts observed in
financial markets.

One of the best known statistical regularities of financial time
series is the fact that the empirical distribution of asset price
changes is fat tailed, i.e. there is a higher probability of extreme
events than in a Gaussian distribution
\cite{Mandelbrot63,Fama64,Officer72,Akigiray89,Mantegna95,Longin96,Lux96,gopi98,Plerou99}. 
Moreover, there are strong indications that the part of the
distribution describing large price changes follows a power-law
\cite{gopi98}. This is important for financial risk, since it means
that large price fluctuations are much more common than one might
expect. There have been several conjectures about the origin of fat
tails in prices \cite{Clark73,Gabaix03}.  Recently \cite{Farmer04} it
has been suggested that fluctuations of liquidity, i.e. the market's
ability to absorb new market orders, could be at the origin of large
price changes. The proposed mechanism for large price change is the
following. Even for the most liquid stocks there can be substantial
gaps in the order book, corresponding to a block of adjacent price
levels containing no quotes. When such a gap exists next to the best
price, a new order can remove the best quote, triggering a large
midpoint price change.  Thus, the distribution of large price changes
merely reflects the distribution of gaps in the limit order book. The
market order triggering the trade must have a size at least equal to
the opposite best and can therefore be of small size.  A market order
producing an immediate large price change also creates a temporary
large spread. The market then reverts the spread to a normal value as
a consequence of the events immediately following the trade. This
paper empirically investigates how the market reacts to these
temporary liquidity crises and how the spread and the price revert
back to normal values.

The presence of large spread poses challenging questions to the
traders on the optimal way to trade. When the spread is large
liquidity takers have a strong disincentive for submitting market
orders given that the cost, the bid-ask spread, is large. Conversely
market makers (liquidity providers) trade by placing limit orders and
therefore profit of a large spread by selling at the ask price and
buying back at the lower bid price. Moreover there is a strong
incentive to place limit orders in the spread given that a trader can
attain the best position (price) in the book with the highest
execution priority. However the optimal order placement inside the
spread is a nontrivial problem. Slowly closing the spread by placing a
limit order at a price just beating the best by one tick and waiting
for a market order would give the best execution price (from the point
of view of the trader placing the limit order), but this strategy
risks being beaten by limit orders of other traders.  After some time
the price reaches a `normal' value attractive to market order
submitters.

Beside the problem of how to close the spread, a large spread also
poses the challenge of establishing the ``right" level of the price
after the temporary liquidity crisis disappears . This problem is
important for all types of traders.  Liquidity providers have to
decide how to readjust their quotes by taking into account the
informed nature of the market order which generated the large spread.
On the other hand for liquidity takers market impact is one of the
most important costs of trading. When a liquidity taker wishes to
submit a large order she usually decides to split it in parts and
trade it incrementally. Any transaction of a part of the large order
pushes the price in a direction that makes the next transaction more
unfavorable for her. Thus liquidity takers wish to minimize the price
change due to their own trading and they need to know what the
permanent part of their own trading is. In the second part of this
paper we investigate empirically the relation between immediate and
permanent impact.

Our paper is organized as follow. In Section II we present the data
set used to perform our empirical analyses. In Section III we present
a graphical representation of the order book that might help to
visualize some aspects of the book dynamics. Section IV presents the
results we have obtained in our investigation of the bid-ask spread
dynamics. The determinants of bid-ask spread decay are discussed in
Section V. In Section VI we investigate the permanent impact both of a
fluctuation and of a transaction altering the spread. In Section VI we
briefly discuss our results and draw some conclusions.

\section{Data}

Our dataset is composed of $71$ highly capitalized stocks traded at
the London Stock Exchange. The time period is the whole year 2002. The
ticker of the investigated stocks are: SHEL, VOD, GSK, RBS, BP., AZN,
LLOY, REL, HSBA, BARC, HBOS, ULVR, BT.A, DGE, AV., PRU, BSY, WPP, RIO,
ANL, TSCO, RTR, PSON, STAN, CBRY, BA., BG., BLT, BATS, NGT, AVZ, CPG,
AAL, ARM, CNA, CW., RSA, KFI, SPW, SUY, IMT, RB., BZP, LGEN, ICI, MKS,
GUS, SSE, DXNS, SHP, ALLD, OOM, BOG, BOC, HG., SCTN, BAA, LOG, RR,
SMIN, HNS, GAA, NYA, SGE, WOS, AL., SFW, ISYS, III, BAY, RTO. The
order of the tick symbol in the list is fixed by its rank when the
stocks are sorted according to the size occupied by the stock in the
database. SHEL occupied the largest amount of memory in the database
while RTO occupied the smallest memory, among the considered stocks.

The LSE has a dual market structures consisting of a centralized limit
order book market and a decentralized bilateral exchange. In London
the centralized limit order book market is called the on-book market
 and the decentralized bilateral exchange is called the off-book
market. In 2002 $62\%$ of transactions of LSE stocks occurred
in the on-book exchange. In our study we consider only the on-book
market. The on-book market is a fully automated electronic
exchange. Market participants have the ability to view the entire
limit order book at any instant, and to place trading orders and have
them entered into the book, executed, or canceled almost
instantaneously. The trading day begins and terminates with an
auction. For this study, we ignore the opening and closing auctions
and analyze only orders placed during the continuous auction period.
Moreover, in order to avoid effects near the start and end of the day,
we omit the first and last half an hour of trading from the
calculation each day. That is we make a time series for each day from
8:30 AM to 4:00 PM and using it we calculate the conditional averages
and the unconditional averages and repeat the process for each
separate day, without including any time lags across different days.
Finally, in most of our analyses, we removed the data of trading
occurring on September 20, 2002. This is because on that day anomalous
behavior of the spread due to unusual trading was observed as
described below.

\section{Grapical Representation of Order Book as Complex Dynamical System}

\begin{figure}[ptb]
\begin{center}
\includegraphics[scale=0.5,angle=-90]{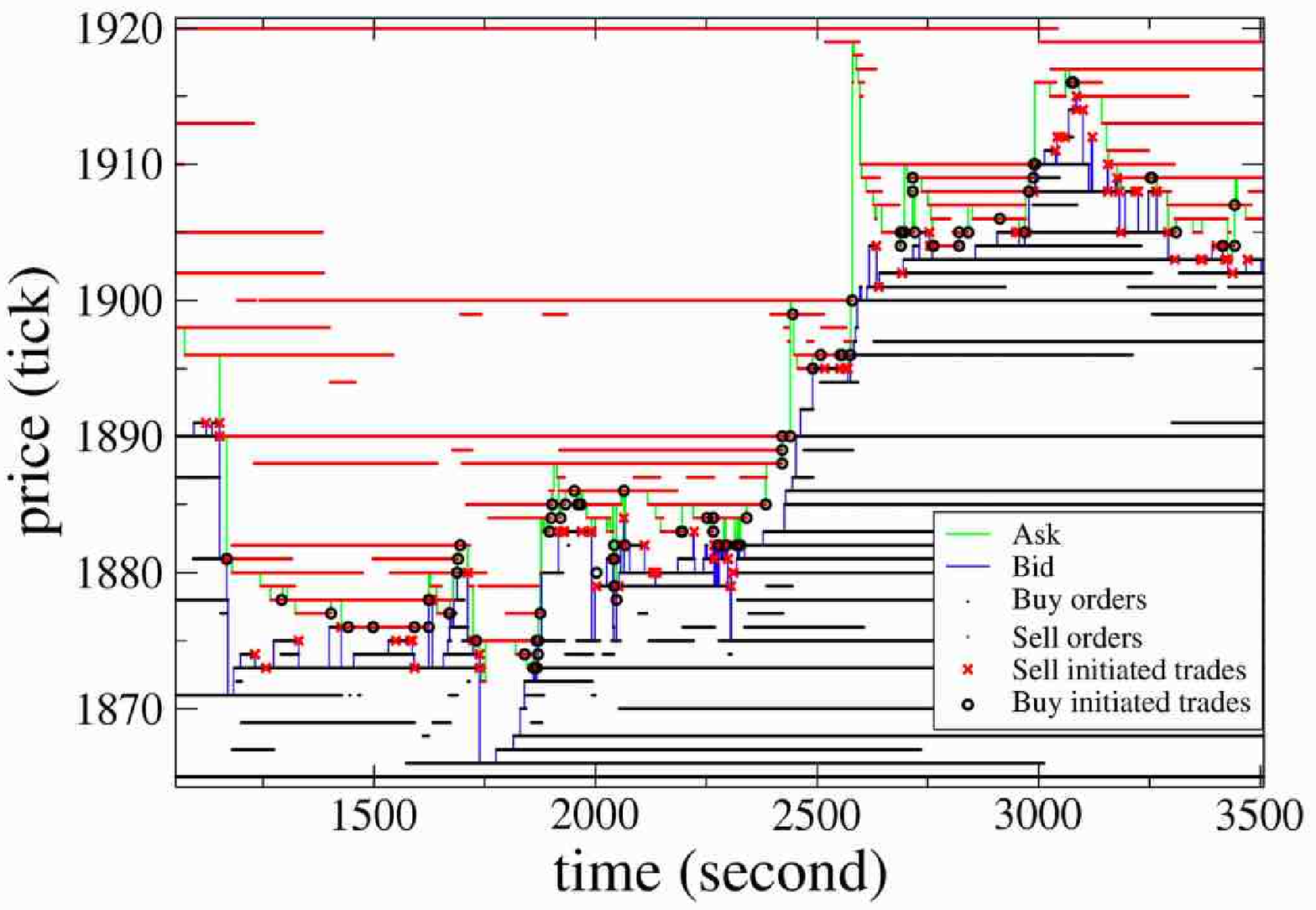}
\includegraphics[scale=0.5,angle=-90]{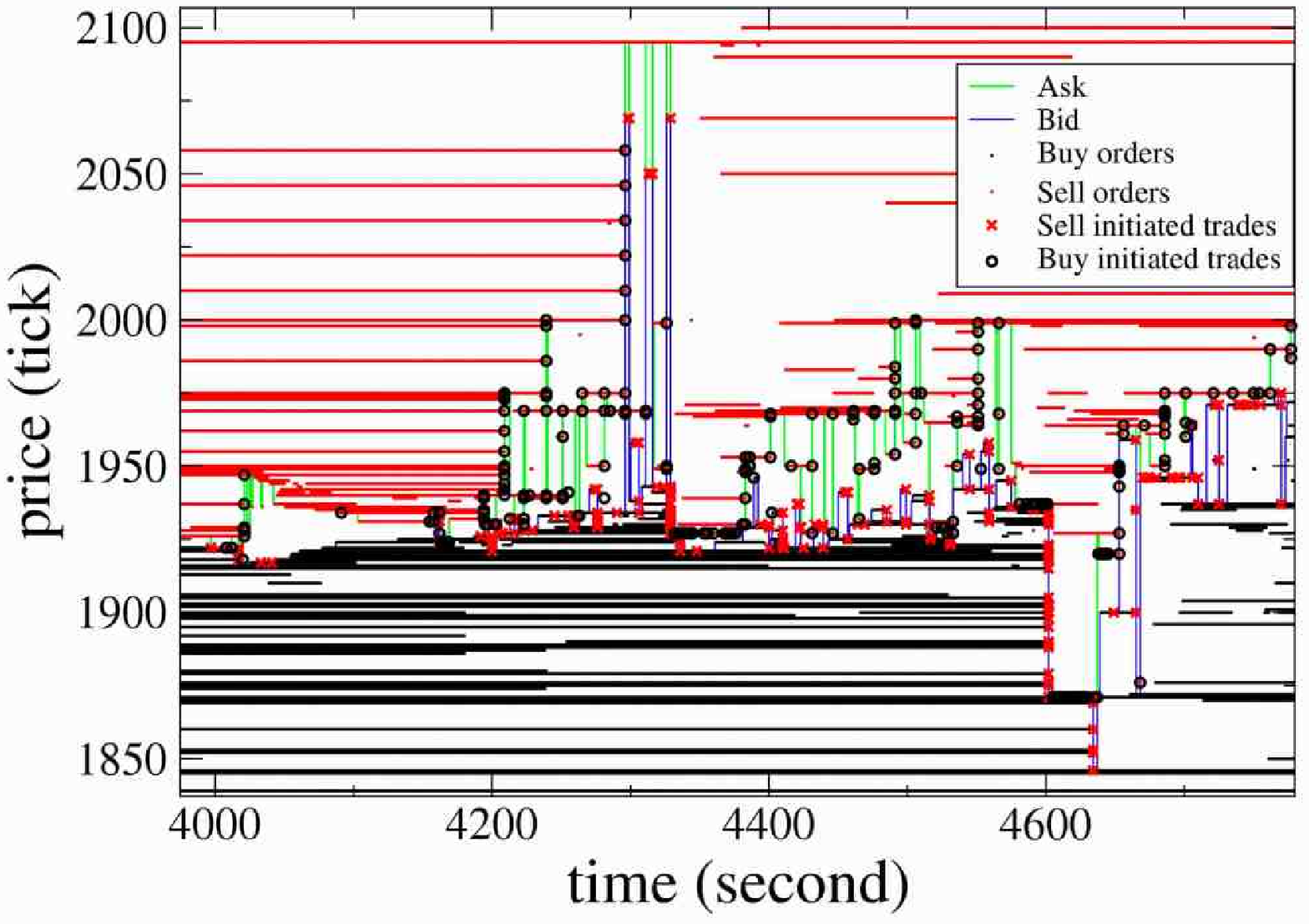}
\end{center}
\caption{Examples of limit order book time series. Top panel: AZN order book dynamics on September 4, 2002 during normal trading. Time is given in seconds from 8:00 am. Bottom panel: AZN order book dynamics on September 20, 2002 when a rogue trading pattern was occurring. Time is given in seconds from 9:00 am.  The plots show all limit order price levels present in the price range shown, as well as the bid (blue line) and ask (green line) and trades (sell initiated - red crosses, buy initiated - black circles).}
\label{AZNts}
\end{figure}

Most financial markets work through a limit order book
mechanism. Agents can place market orders, which are requests to buy
or sell a given number of shares immediately at the best available
price or limit orders which also state a limit price, corresponding to
the worst allowable price for the transaction.  Limit orders often
fail to result in an immediate transaction, and are stored in a queue
called the \textit{limit order book}. At any given time there is a
best (lowest) offer to sell with price $a(t)$, and a best (highest)
bid to buy with price $b(t)$.  These are also called the {\it ask} and
the {\it bid}, respectively.  The difference $s(t)=a(t)-b(t)$ between
the ask and the bid is called the \textit{spread}. The {\it midprice}
is defined as $m(t)=(a(t)+b(t))/2$. The difference between the best
buy price and the second best buy price is called the first buy gap,
whereas the difference between the second best sell price and the best
sell price is called the first sell gap. Gaps provide a proxy for the
immediate liquidity present in the limit order book.

Visualizing the dynamics of the limit order book is a complex task,
because many orders are present in the book at a given time. We
represent the book dynamics with a graph of the type shown in
Fig.~\ref{AZNts} for the stock Astrazeneca (AZN) in two representative
days. The top panel shows an example of a normal trading period
recorded on September 4, 2002, whereas the bottom panel shows an
unusual day, specifically a period of September 20, 2002 when an
unusual rogue trading pattern was occurring.  In these plots each
line shows a price level. Price levels appear as the result of orders
being submitted into the book. Similarly price levels may disappear
due to cancellations of orders or due to trades. Of course there may
be other orders submitted onto existing price levels, but these are
not explicitly shown in the plots. The ask is shown as a green line
and the bid is shown as a blue line. The first sell gap is the block
of unoccupied price levels above the ask before the next sell price
level and the first buy gap is the block of unoccupied price levels
below the bid before the next buy price level.  Indeed the difference
between the two figures, although both are AZN just a few trading days
apart illustrates the heterogeneity of order book dynamics.

The September 4 trading is normal and representative of other trading
days. Here price drifting is visible by a tick by tick order
submission process, as well as some large fluctuations. Large
fluctuations occur when there is a large first gap and a trade (or
sometimes cancellation) removes all the quotes at the best price, such
as can be seen around $t\approx 2500$. This large fluctuation creates
a large spread. The spread relaxes generally with a slow dynamics to a
more normal value, in part due to tick by tick `price beating' order
submissions into the spread.  Such price beating can occur on the same
side of the book as the trade which removed the best and created the
large spread, in which case the action is to revert the
midprice. Alternatively it can occur on the opposite side, in which
case the action is to produce midprice drifting.

In the example from September 20 many exceptional aggressive market
orders are submitted. These orders cannot be filled solely by orders at
the best so they cut across several price levels in the book, creating
a highly volatile spread dynamics. The spread can become huge, of the
order of a hundred ticks, and large midprice fluctuations result. It
should be noted that the scale of price axis is quite different in the
two panels of Fig.~\ref{AZNts}. In fact in the top panel of
Fig.~\ref{AZNts} the y axis covers slightly more than 50 ticks whereas
the same axis cover more than 150 ticks in the bottom panel.

The order book dynamics presents fluctuations in order submission
rates, cancellation rates and trade rates which depend on the spread
and size of preceding price fluctuations. All these fluctuations
produce a non-trivial price-time coupling and `cascade' like dynamics.
When the order book is plotted as in Fig.~\ref{AZNts} some complex
patterns of order book dynamics become evident. One particular example
is the time asymmetry created by the spread dynamics, whereby the
spread opens by few large fluctuations and closes by many small
ones. When one studies the midprice or return time series this time
asymmetry is not apparent as in the direct visualization obtained by
the kind of plot presented in Fig.~\ref{AZNts}. This type of plot is
an extension of the plots originally presented in
\cite{Biais95}. However, differently from previous plots, the present
version contains full information about the status of the order
book. This kind of figure can be a useful tool for investigation of
the order book dynamics during days when anomalous trading behavior is
present. Moreover, a direct investigation of the bottom panel of
Fig.~\ref{AZNts} shows that this graphical tool can also be useful for
distinguishing different types of high-frequency trading patterns.

\section{Spread Analysis}

\begin{figure}[ptb]
\begin{center}
\includegraphics[scale=0.33,angle=-90]{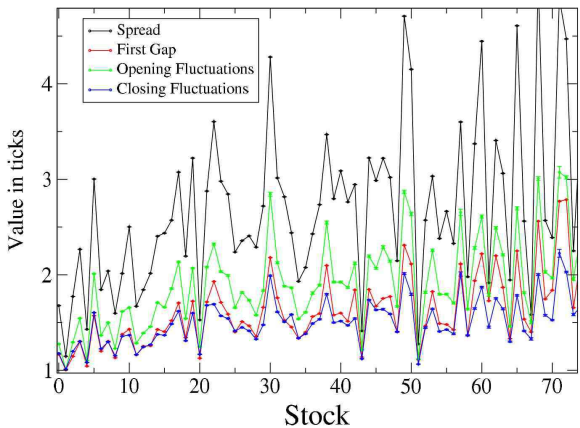}
\end{center}
\caption{
Expected value of spread, first gap, positive and negative spread fluctuations for $71$ stocks in our database. All the quantities are expressed in ticks and the stocks are ordered by size of database. For the definition of these quantities see the text.}
\label{0123comp}
\end{figure}

The time series describing the dynamics of the spread is characterized
by two stylized facts (see, for example, \cite{Plerou05}). First, the
unconditional spread distribution has a density function with a fat
tail. Some papers suggest that the tail of the spread distribution is
well fitted by a power law function \cite{Plerou05,Mike06} whereas in
other studies the spread seems to have an exponential tail
\cite{Bouchaud06}. The second fact is that spread seems to be
described by a long memory process. This implies that the
autocorrelation of the spread decays asymptotically as a power law
with an exponent smaller than one.  Beside these facts there must be a
relation between the spread and the variables determining the spread
dynamics. These variables are the gap size and the spread
variation. The spread is a mean reverting process. Market orders and
cancellations at the best can increase the spread whereas limit orders
in the spread decrease the spread.  An overview of how the spread and
related quantities varies across the first 71 stocks in our database
is shown in Figure~\ref{0123comp}. All quantities are sampled every
second. We compute the expected spread $E(s(t))$, the expected
(symmetrized) first gap $(1/2)E(a_2(t)-a(t))+(1/2)E(b(t)-b_2(t))$,
where $a_2(t)$ and $b_2(t)$ denote the second ask price level and
second bid price level respectively. Here and in the following
$E(...)$ indicates an average over time $t$. Moreover, for spread
opening fluctuations and spread closing fluctuations, the expectation
is taken given there is a fluctuation in either the bid or ask or
both. Opening fluctuations consist of the average of
$E(a(t+1)-a(t)|a(t+1)>a(t))$ and $E(b(t)-b(t+1)|b(t+1)<b(t))$ with
analogous expressions for the closing fluctuations.  Finally in the
figure the stocks are ordered by size of database. The database size
roughly corresponds with activity occurring in each stock during the
year 2002, where high activity means a high order submission rate, a
high trading rate etc. This therefore suggests that in general the
spread, and spread related quantities, decreases with increasing
activity.  The expected size of spread opening fluctuations is
strongly related to the expected size of the first gap when an event
removes all the orders at the best price. Similarly the expected size
of spread closing fluctuations is strongly related to the expected
size of the first gap created by an order submitted into the
spread. Fig.~\ref{0123comp} shows that spread closing fluctuations are
smaller in size than spread opening fluctuations. This means spread
closing fluctuations must be more numerous than spread opening
fluctuations to maintain a stationary spread. This is therefore a
consequence of the slow decay of the spread as will be described
below. Since midpoint fluctuations are made up of both opening and
closing fluctuations, this figure also suggests that midpoint
volatility increases with spread and therefore decreases with
activity.

\begin{figure}[ptb]
\begin{center}
\includegraphics[scale=0.33,angle=-90]{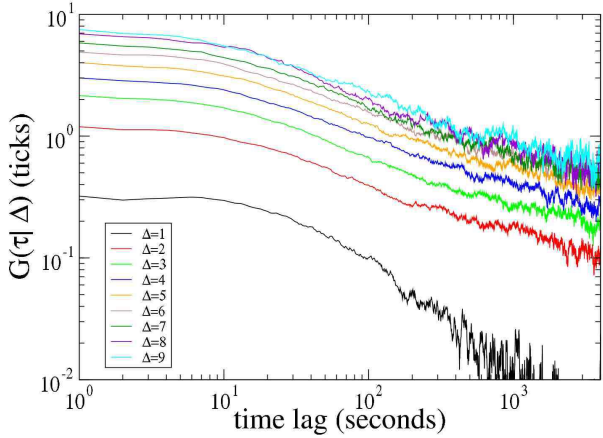}
\includegraphics[scale=0.33,angle=-90]{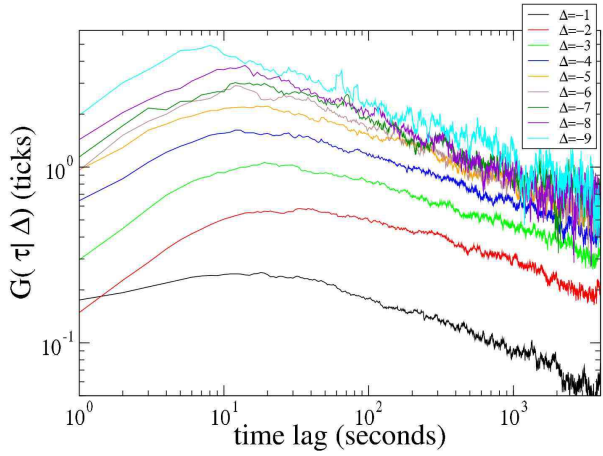}
\end{center}
\caption{Conditional spread decay $G(\tau|\Delta)$ defined in
Eq.~\ref{g} for the stock AZN. Top panel shows $G(\tau|\Delta)$ for
different positive values of $\Delta$ (in ticks) corresponding to an
opening of the spread at time lag $\tau=0$. Bottom panel shows
$G(\tau|\Delta)$ for different negative values of $\Delta$ (in ticks)
corresponding to a closing of the spread at time lag $\tau=0$.}
\label{AZNcondspread}
\end{figure}

\begin{figure}[ptb]
\begin{center}
\includegraphics[scale=0.33,angle=-90]{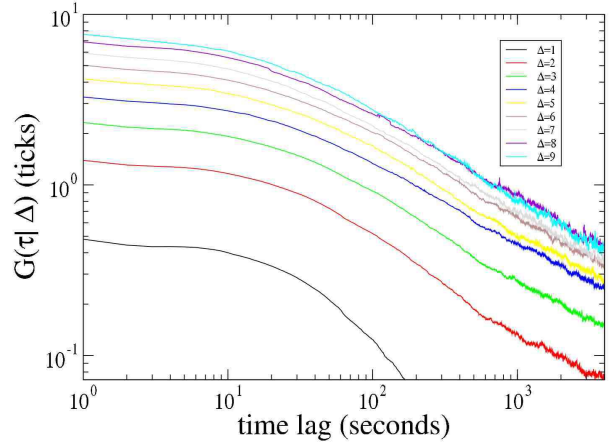}
\includegraphics[scale=0.33,angle=-90]{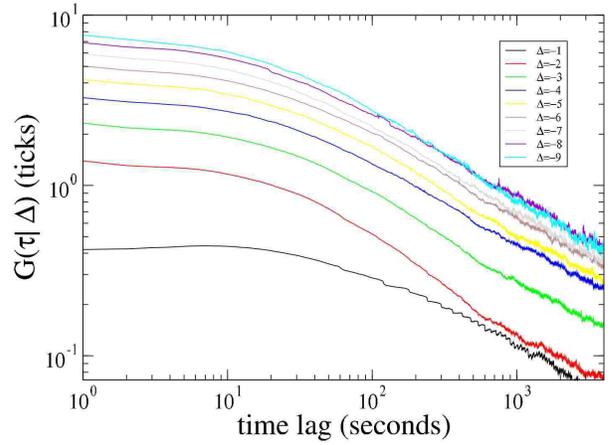}
\end{center}
\caption{Conditional spread decay $G(\tau|\Delta)$ defined in
Eq.~\ref{g}. The curves are obtained by averaging $G(\tau|\Delta)$
over the 71 stocks of our sample. Top panel shows $G(\tau|\Delta)$ for
different positive values of $\Delta$ (in ticks) corresponding to a
opening of the spread at time lag $\tau=0$. Bottom panel shows
$G(\tau|\Delta)$ for different negative values of $\Delta$ (in ticks)
corresponding to a closing of the spread at time lag $\tau=0$.}
\label{poolcondspread}
\end{figure}

In this paper we are mainly interested in the conditional dynamics of
spread. We wish to characterize the dynamics of the spread $s(t)$
conditional to a spread {\it variation}. In other words, we wish to
answer the question: how does the spread return to a normal value
after a spread variation?  To this end we compute the quantity
\begin{equation}
G(\tau|\Delta)=E(s(t+\tau)|s(t)-s(t-1)=\Delta)-E(s(t))
\label{g}
\end{equation}
Figure~\ref{AZNcondspread} shows this quantity for the stock AZN as a
function of $\tau$ for different positive and negative values of
$\Delta$. The decay of $G(\tau|\Delta)$ as a function of $\tau$ is
very slow and for large values of $\tau$ is compatible with a power
law decay. In order to obtain better statistics in
Figure~\ref{poolcondspread} we plot $G(\tau|\Delta)$ averaged over the
$71$ stocks of our sample\footnote{We are aware that aggregating data
from different stocks can create biases and/or spurious statistical
effects in the estimation process. However the comparison of the
averaged data with the data from different individual stocks suggests
that the power law decay of the spread is a common behavior to many
stocks.}. As in the individual case the asymptotic decay is compatible
with a power law, $G(\tau|\Delta)\sim \tau^{-\delta}$, and the fitted
exponent $\delta$ is around $0.4-0.5$.  By comparing $G(\tau|\Delta)$
for positive and negative values of $\Delta$ we notice that spread
decay conditional to a positive value $\Delta>0$ is very close to the
spread decay conditional to the negative value $-\Delta-1$ for time
lags longer than few hundreds seconds. We do not have an explanation
for this empirical observation.  The slow decay of the spread
indicates that large changes in the spread are reverted to a normal
value with a very slow dynamics. The power law fit suggests that there
is not a typical scale for the spread decay. A similar slow decay of
the spread was recently observed by Zawadowski {\it et al.} in
Refs.~\cite{Kertesz1,Kertesz2}. The main difference with this work is
that Zawadowski {\it et al.} investigated the decay of the spread
conditional on a negative change in the midprice rather than in the
spread itself. Moreover, the investigated market and database is quite
different. Zawadowski {\it et al} investigated the NYSE and the NASDAQ
by using the Trade and Quote (TAQ) database. They found a slow decay
of the spread at NYSE but not at NASDAQ. This does not seem to be
consistent with our findings especially when considering the fact that
the LSE is probably closer to NASDAQ than to NYSE due to the presence
of the specialist at NYSE.

It is worth noting that the slow spread decay is not a consequence of the
long memory property of the spread itself. To show why this is the
case, let us consider a generic zero mean stochastic process $x(t)$. The
quantity $G(\tau|\Delta)$ is the expectation value
$E(x(t+\tau)|x(t)-x(t-1)=y(t)=\Delta)$. In general
\begin{equation}
E(x(t+\tau)y(t))=\int~y(t) E(x(t+\tau)|y(t)) p(y)~dy
\end{equation}
thus the $\tau$ dependence of $E(x(t+\tau) y(t))$ is the same as the
$\tau$ dependence of $E(x(t+\tau)|y(t))$. For analytical convenience
we compute the $\tau$ dependence of $E[x(t+\tau)~(x(t)-x(t-1))]$
rather than the conditional expectation of Eq.~\ref{g}.  The quantity
$E[x(t+\tau)~(x(t)-x(t-1))]$ is equal to $\rho(\tau)-\rho(\tau-1)$,
where $\rho(\tau)=E(x(t+\tau)x(t))$ is the autocovariance of
$x(t)$. Suppose that $x(t)$ is a long memory process, i.e. that
$\rho(\tau)\sim A\tau^{-\beta}$ with $0<\beta<1$. Then our argument
shows that
\begin{equation}
E[x(t+\tau)~(x(t)-x(t-1))]\sim\frac{A\beta}{\tau^{\beta+1}}.
\end{equation}
This result shows that by assuming the spread to be a long memory
process with an autocorrelation function decaying as $\tau^{-\beta}$
with $\beta<1$, one should expect that $G(\tau|\Delta)$ decays
asymptotically as a power law but with an exponent $1+\beta$ larger
than $1$. The fact that the exponent $\delta$ describing the
asymptotic decay of $G(\tau|\Delta)$ is significantly smaller than $1$
indicates that the observed slow decay of $G(\tau|\Delta)$ is not a
consequence of the long memory property of the spread.

\section{Determinants of the spread decay}

What is the order placement process giving rise to the slow decay of
the spread observed in the previous section? The answer to this
question is complicated due to the different types of processes that
contribute to the spread dynamics. In this section we describe the
statistical properties of the events that could contribute to the slow
decay of the spread. This approach does not give a full explanation
for the decay, but highlights the elements that are relevant for the
process.

\begin{figure}[ptb]
  \begin{center}
\includegraphics[angle=-90,scale=0.33]{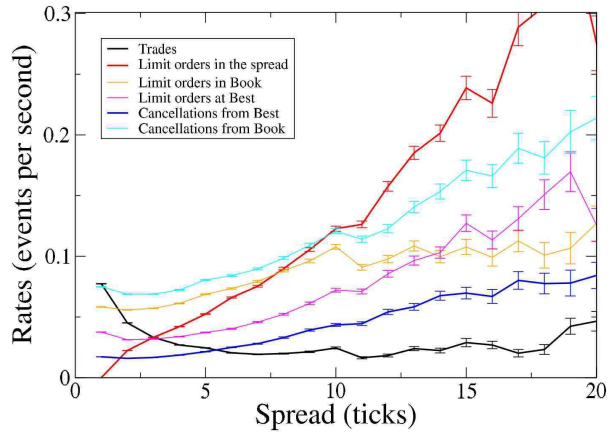}
\caption{Rates of a series of events in the limit order book conditional
to the size of the spread. Limit orders are divided into limit orders
in the spread, limit orders at an existing best (bid or ask) and limit
orders placed inside the book. Similarly cancellations are divided
into cancellations of limit orders at the best and cancellations of
limit orders inside the book (at the time when they are cancelled). The
data refer to the stock AZN.}
\label{ratescondspread} 
\end{center}
\end{figure}

The dynamics of the spread is determined by the flow of market orders,
limit orders falling in the spread and cancellations of the orders at
the best bid and ask. The rates of the three different types of orders
strongly depend on the spread size. Figure~\ref{ratescondspread} shows
the rates (in events per second) of different possible events in the
limit order book, specifically trades (market orders), limit orders
and cancellations. Limit orders are divided in three subsets according
to their limit price. We consider limit orders in the spread, limit
orders at an existing best (bid or ask) and limit orders placed inside
the book. Similarly cancellations are divided into cancellations of
limit orders at the best and cancellations of limit orders inside the
book (at the time when they are cancelled).
Figure~\ref{ratescondspread} shows that the rate of trading decreases
as the spread increases, whereas the rate of limit order submission in
the spread dramatically increases with spread size.  A similar
behavior of the rates of order submission conditional to the spread
size has been recently observed by Mike and Farmer \cite{Mike06}.
This behavior is expected since a large bid-ask spread is a strong
disincentive to trade, given that the spread related cost is large. On
the other hand, a large spread is an incentive for limit order
placement inside the spread in order to have priority of execution at
a convenient price. Also the cancellations rate increases with
spread. These findings are consistent with a process whereby liquidity
providers cancel and replace limit orders in order to slowly close the
spread. The increase of the rate of limit order and cancellation at
the best and the decrease of market order rate, are consistent with
the view of the slow decay of the spread observed in the previous
section.

\begin{figure}[ptb]
  \begin{center}
\includegraphics[angle=-90,scale=0.33]{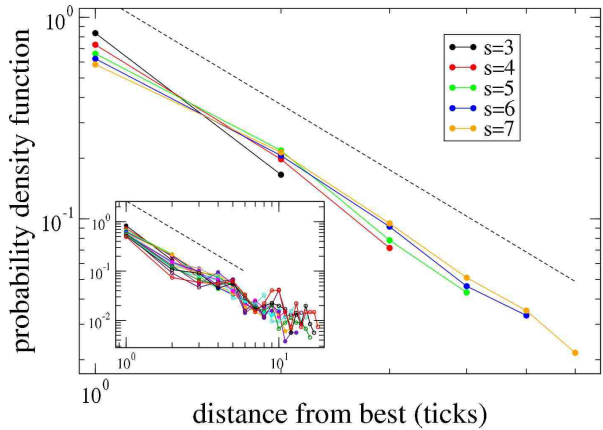}
\caption{Distribution of the distance between price of limit order in
the spread and the best available price (ask for sell limit orders and
bid for buy limit orders), for various values of the spread $s$. The
inset shows more values of the spread size $s$. The data refer to the
stock AZN.}
\label{LOinspread} 
\end{center}
\end{figure}

The way in which limit orders are placed into the spread when the
spread is large is another determinant of the dynamics of spread
decay. Limit order placement in the spread follows an interesting
scaling relation observed originally by Mike and Farmer
\cite{Mike06}. In Figure~\ref{LOinspread} we show the distribution of
the distance from the same best of limit orders placed in the spread
for different values of the spread.  Specifically, given a spread size
$s(t)=a(t)-b(t)$ and a limit order with price $p$ between the ask
$a(t)$ and the sell $b(t)$, we consider the distribution of $a(t)-p$
for sell limit orders and $p-b(t)$ for buy limit orders. The shape of
the curves shown in Figure~\ref{LOinspread} is consistent with a power
law decay with an exponent $\sim 1.8$. Mike and Farmer \cite{Mike06}
fits the limit order placement with a Student distribution with $1.3$
degrees of freedom. This distribution indicates that when the spread
is large, limit orders are not placed simply in a way that immediately
reverts the spread back to its typical (small) value. Rather limit
orders are sequentially placed close to the existing best price and
this leads to a slow decay of the spread.

\begin{figure}[ptb]
  \begin{center}
\includegraphics[angle=-90,scale=0.33]{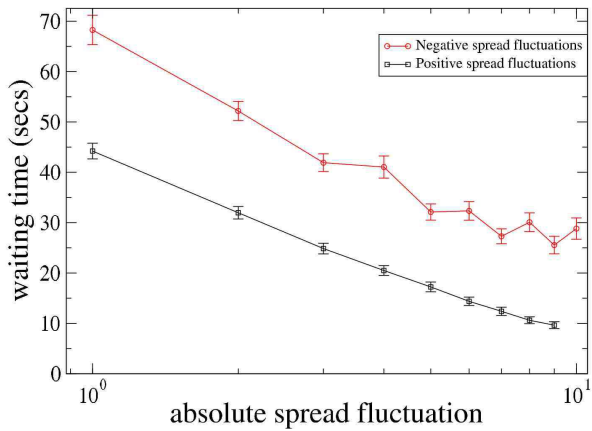}
\caption{Mean waiting time between a spread variation
$\Delta=s(t)-s(t-1)$ and the next spread changing event as a function
of $|\Delta|$ for positive (black symbols) and negative (red symbols)
spread fluctuations. The figure shows an average of the mean waiting
time across the stocks of our sample.}
\label{waitingtime} 
\end{center}
\end{figure}

In addition to the investigation of the rate of orders as a function
of the size of the spread (Fig.~\ref{ratescondspread}) one can also
measure the time interval between a spread variation of a given size
and the next spread variation (of any size). This waiting time gives
the reaction time of the market to an abrupt spread variation. In
Figure~\ref{waitingtime} we show the mean waiting time between a
spread variation $\Delta= s(t)-s(t-1)$ and the next spread variation
as a function of $\Delta$. The waiting time decreases when the spread
variation increases and the functional dependence is approximately
logarithmic. In other words the larger the spread variation, the
shorter the time one has to wait until a new event changes the spread
again. Moreover the waiting time for positive spread variations is
much shorter than the waiting time for negative spread variations of
the same size (in absolute value). This result is to be expected and
shows that the market reacts faster to an increase in the spread
rather than to a decrease in the spread. As a last remark on the mean
waiting time, we note that there is an oscillation like behavior in
the negative spread variation, but we do not have an explanation for
this observation.

\begin{figure}[ptb]
  \begin{center}
\includegraphics[angle=-90,scale=0.33]{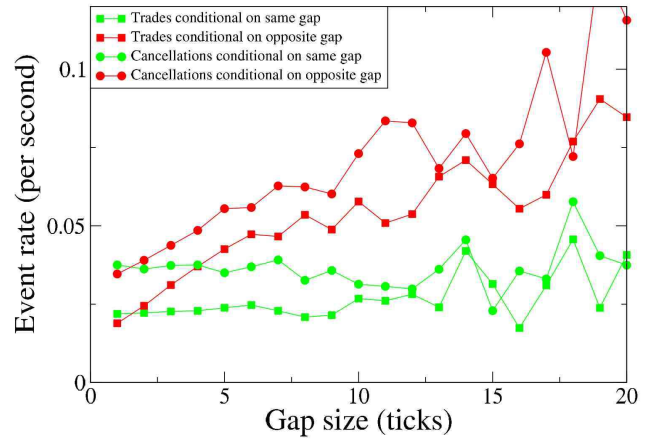}
\caption{Rate of trades (squares) and cancellations (circles)
conditional to the size of the same (green) or opposite (red) gap. The
data refer to the stock AZN.}  
\label{condgap} 
\end{center}
\end{figure}

Beside the spread and its fluctuations another important quantity
determining liquidity is the gap size. As described above the gap size
is the absolute price difference between the best available price
(e.g. to buy) and the next best available price. Gap size is important
because it has been suggested that immediate price impact is strongly
determined by gap size \cite{Farmer04}.  In any given instant there
are two gaps, one on the buy side and one on the sell side of the
limit order book. For a buy (sell) market order we define the same
side gap size as the gap size on the buy (sell) limit order book side,
and opposite side gap size as the gap size on the sell (buy) limit
order book side. In Figure~\ref{condgap} we show the trade rate
conditional to the same and to the opposite gap size. We see that
while the trade rate is almost independent of same side gap size, the
rate increases significantly with the opposite side gap size. A
possible interpretation of this result is the following. Imagine the
market is drifting upward - i.e. the price is increasing. Then most
trades are buy initiated and the gap on the sell side is large. Buy
limit orders tend to be submitted just inside the spread i.e. beating
the best buy by 1 tick, so the gap on the buy side is small.

A similar result is seen for cancellations. Figure~\ref{condgap} also
shows the cancellation rate conditional to the size of the same and of
the opposite gap. Again it is seen that the cancellation rate weakly
depends on the same side gap size, whereas it increases with the size
of the opposite gap. When the price is drifting, for example upward,
there is a strong limit order flow and cancellation on the buy side of
the book. As described above this might be due to the fact that
liquidity providers try to gain the best bid by placing buy limit
orders in the spread and canceling beaten buy limit orders to get a
better position.

\section{Permanent impact}

In this section we consider the permanent impact of a price
fluctuation. The bid or ask can fluctuate in several ways. Firstly a
limit order can be submitted into the spread, secondly a cancellation
can remove the best, and thirdly a trade can remove the best. The
second possibility is rather rare because there are usually several
orders at the best owned by different trading agents and they all have
to be cancelled independently. On the other hand a single submitted
market order can remove all the orders at the best price in one trade.
Indeed when a market order arrives in the market, it may trigger a
trade which creates a price change.
This immediate price change is the immediate impact. The properties of
immediate impact as a function of the trading volume and of the market
capitalization of the stock have been studied for example
in~\cite{Lillo03,Bouchaud03,lillocomment}. The transaction and the
consequent price change generate a cascade of events in
reaction. After a sufficiently large period of time the effect of the
trade has vanished and the price will be in general different from the
price before the trade. The variation of price is the permanent
component of the impact of the trade. In this section we are
interested in measuring this permanent impact. Since price can
fluctuate for different reasons, in the following we distinguish
between {\it fluctuation impact} and {\it transaction impact}. With
the first term we refer to the impact on the price conditional to a
price fluctuation (caused by any type of event) happening at a
previous time. Instead transaction impact is the impact on the price
conditional to a fluctuation in the price and to the presence of (at
least) a trade at a previous time. Moreover, since there are
different prices in the market at a given time (bid, ask, midprice,
etc.), we consider the impact on the ask and on the midprice. By
considering the ask price fluctuations we can separate the different
effects of trades causing positive fluctuations and limit orders
falling in the spread causing negative fluctuations

\subsection{Fluctuation Impact}

\begin{figure}[ptb]
  \begin{center}
\includegraphics[scale=0.33,angle=-90]{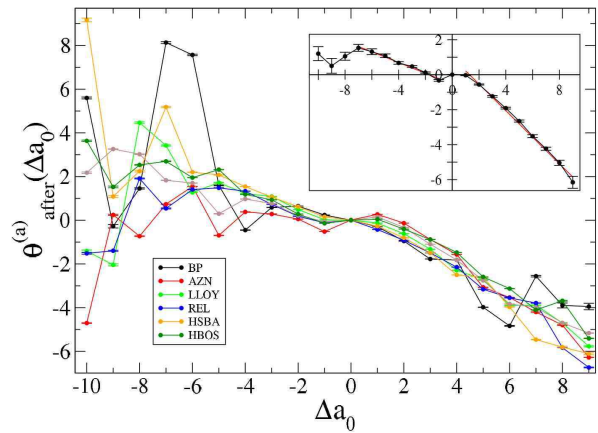} 
  \caption{Ask impact
$\theta^{(a)}_{after}(\Delta a_0)$ for 6 highly traded and representative stocks
versus $\Delta a_0$. The error bars are the
standard errors. The inset shows the ask permanent impact averaged across the
first 55 stocks of our database, and here the error bars show the stock ensemble
average standard errors. Linear regression fits to the points $\Delta
a_0=1,...,9$ and $\Delta a_0=-1,...,-7$ are shown as red lines. For
the positive $\Delta a_0$ range we find 
$\theta^{(a)}_{after}(\Delta a_0)=0.98-0.76\Delta a_0$ while for the negative range we obtain
$\theta^{(a)}_{after}(\Delta a_0)=-0.53-0.3\Delta a_0$.}
  \label{permimpb}
  \end{center}
\end{figure}

As mentioned above fluctuation impact is the impact on the price
conditional to a price fluctuation at a previous time.  We consider
first the permanent impact of ask price fluctuations.  Consider the
events happening at time $t$. The ask price changes due to these
events by a quantity $\Delta a_0=a(t)-a(t-1)$, where $a(t)$ is the ask
price at the end of second $t$. $\Delta a_0$ is the immediate impact.
After a sufficiently large time lag $\tau$ the ask price will be
$a(t+\tau)$ and the permanent impact is $\Delta
a_{perm}(\tau)=a(t+\tau)-a(t-1)$. Thus the permanent impact can be
decomposed as
\begin{equation}
\Delta a_{perm}(\tau)=\Delta a_0+\Delta a_{after}(\tau)
\end{equation}
where $\Delta a_{after}(\tau)=a(t+\tau)-a(t)$ is the price change due
to order submission and other events happening after the trade at time $t$ has been
completed. $\Delta a_{after}$ is the reaction of the market to the
event at time $t$.  We measure first the conditional quantity
\begin{eqnarray}
E(\Delta a_{after}(\tau)| \Delta a_0) = \\
E[a(t+\tau)-a(t)|\Delta a_0]-E[a(t+\tau)-a(t)]. \nonumber
\end{eqnarray}
We subtract the unconditional mean $E[a(t+\tau)-a(t)]$ in order to
avoid spurious effects due to the finiteness of the time series.
If we let $\tau \to \infty $ then we obtain the permanent impact. Since we cannot take the limit $\tau
\to \infty$ in the calculation of the permanent impact, we use as a proxy
\begin{equation}
\theta^{(a)}_{after}(\Delta a_0)=\frac{1}{T}\sum_{\tau=t_1}^{t_1+T}E(\Delta a_{after}(\tau)| \Delta a_0)
\end{equation}
where we take $t_1=500$ seconds and $T=500$ seconds. In Fig.\ref{permimpb} we show $\theta^{(a)}_{after}(\Delta a_0)$ for some highly traded and representative stocks. We also show the ask impact averaged across the first 55 stocks of our list in the inset presented in the figure. 

If the ask price time series is a martingale we expect $\theta^{(a)}_{after}(\Delta a_0)=0$ independent of $\Delta a_0$. That this does not hold is
shown in Fig.\ref{permimpb}. There are large variations across stocks,
some showing trend following for small positive $\Delta a_0$ and all
showing partial reversion for large positive $\Delta a_0$. 
As described above, positive $\Delta a_0$ correspond to trade (or cancellation) initiated
fluctuations. We note that a one tick positive immediate fluctuation in
AZN $\Delta a_0=1$, induces on average a further 0.25 tick positive
fluctuation in the long time average $\theta^{(a)}_{after}(\Delta a_0)$.
This shows the presence of a trending phase of the ask price in some
stocks, which reinforces the direction of the price change. In other stocks, for example BP, a one tick positive ask change induces a negative fluctuation of $\theta^{(a)}_{after}(\Delta a_0)$ 
The inset of Fig.~\ref{permimpb} shows the average behavior of $\theta^{(a)}_{after}(\Delta a_0)$ across $55$ stocks. 
A clear linear behavior of  $\theta^{(a)}_{after}(\Delta a_0)$ as a function of $\Delta a_0$ can be seen in different $\Delta a_0$ intervals. Also there is a significant asymmetry in $\theta^{(a)}_{after}(\Delta a_0)$ for positive and
negative $\Delta a_0$ since their permanent impact behavior is quite
different. 
In particular for the range $\Delta a_0=2,...,9$ ticks
the points lie on a line of slope approximately $-3/4$, while the
$\Delta a_0=1$ point is close to zero. In other words, large positive spread
fluctuations are partially reverted, while positive one tick spread
opening fluctuations are balanced on the long run. The reverting
behavior after a large positive ask fluctuation is related to the
decay of the large spread opened up by the fluctuation discussed
above. This shows that positive ask fluctuations have both a $\Delta a_0$
independent part and a part which depends on $\Delta a_0$. The total
fluctuation composed of the initial fluctuation and of the successive part
is $\theta^{(a)}_{perm}(\Delta a_0) \sim \alpha_a + \beta_a \Delta a_0$, where
$\alpha_a$ is roughly one tick and $\beta_a$ is approximately $1/4$ ticks.
Negative ask fluctuations,  $\Delta a_0=-1$,
i.e. orders just beating the ask by one tick, are likely
to be followed by further sell orders falling in the spread since they
have a trend following permanent impact of approximately $1/2$
tick. This is again the price beating behavior described above.
Larger spread closing fluctuations, negative $\Delta
a_0$, show a negative slope for the range $-2,...-7$. This means that
spread closing fluctuations in this range are themselves
reverted. This may be because the equilibrium spread is in general not
the possible minimum one tick, but maybe several ticks, and smaller
spreads than equilibrium may be created by orders falling into the
spread which must then revert to equilibrium.

\begin{figure}[ptb]
  \begin{center}
\includegraphics[angle=-90,scale=0.33]{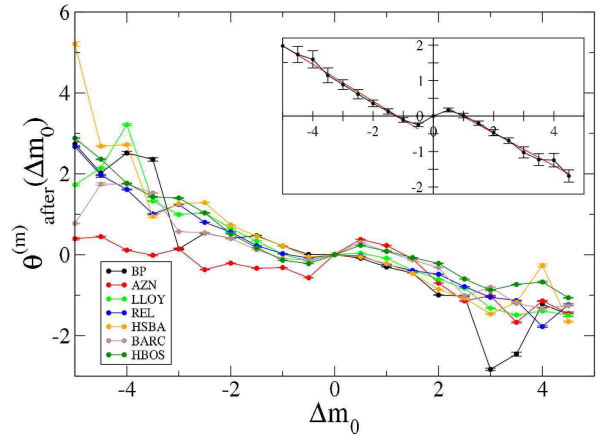} 
  \caption{Midprice impact $\theta^{(m)}_{after}(\Delta m_0)$. 
The figure shows the result obtained for 6 highly capitalized stocks and the error bars denote standard errors. The inset shows $\theta^{(m)}_{after}(\Delta m_0)$ averaged across
$55$ stocks. In this case the error bars show the stock
ensemble average standard errors. Red lines are linear fits.}
  \label{permimpa}
  \end{center}
\end{figure}

\begin{figure}[ptb]
  \begin{center}
\includegraphics[angle=-90,scale=0.33]{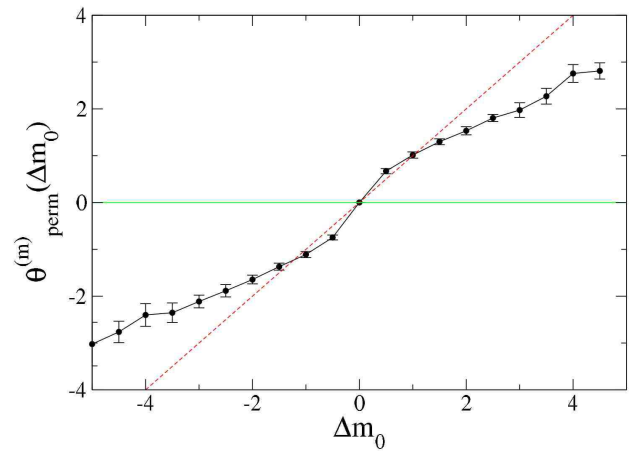} 
  \caption{Midprice permanent impact $\theta^{(m)}_{perm}(\Delta m_0)$
averaged across the first 55 stocks. The error bars show the stock
ensemble average standard errors. The red line is the line $y=x$ that would be obtained in case of a permanent impact equal to the immediate impact $\Delta m_0$. The green line is the line $y=0$ that would obtained in case of zero permanent impact, i.e. complete reversion.}
  \label{trueperm}
  \end{center}
\end{figure}

We now consider the fluctuation impact on the midprice. 
Midprice permanent impact is defined analogously to the ask price permanent impact as
$E(\Delta m_{after}(\tau)| \Delta m_0) = E[m(t+\tau))-m(t)|\Delta m_0] -
E[m(t+\tau)-m(t)]$ where $m(t)$ is the midprice at time t and $\Delta
m_0=m(t)-m(t-1)$ is the mid price one second immediate impact. Again
we let $\tau \to \infty $ to obtain the permanent impact 
\begin{equation}
\theta^{(m)}_{after}(\Delta m_0)=\frac{1}{T}\sum_{\tau=t_1}^{t_1+T}E(\Delta m_{after}(\tau)| \Delta m_0)
\end{equation}
In Fig.\ref{permimpa} we
show the average of $\theta^{(m)}_{after}(\Delta m_0)$. 
As for the ask $\theta^{(m)}_{after}(\Delta m_0)$ is significantly different from zero independently of $\Delta m_0$. Some stocks show quite strong trend following effects for small values of $\Delta m_0$ with quite
strong reversion effects for large values of $\Delta m_0$. For example
a half tick positive immediate fluctuation in AZN, $\Delta m_0=1/2$,
induces on average a further half tick positive fluctuation in the
long time average, $\theta^{(m)}_{after}(\Delta m_0)$ and furthermore a half tick
negative immediate fluctuation can induce a further half tick negative
permanent change. On the other hand, larger immediate fluctuations are
followed by changes of the opposite sign, i.e. partial
reversion. There are however quite strong variations across stocks -
for example HBOS does not seem to show the strong trend following
behavior for small $\Delta m_0$ seen in AZN and there is also varying
degrees of asymmetry between positive and negative $\Delta m_0$
fluctuations.
As seen in the inset in Fig.\ref{permimpa}, the average over the first
55 stocks shows again a clear piecewise linear behavior. When $\Delta m_0=1/2$
($\Delta m_0=-1/2$) tick the midprice is on average $0.17$
($-0.25$) tick. This again shows the presence of a trending phase of
the midprice, which reinforces the direction of the price change. For
larger price change there is on the contrary a partial reversion of
the price. For positive fluctuations the $\Delta m_0$ a linear fit gives the relation
$\theta^{(m)}_{after}(\Delta m_0)=0.45-0.46\Delta m_0$ while for the negative range we
obtain $\theta^{(m)}_{after}(\Delta m_0)=-0.62-0.52\Delta m_0$. The independent part
is $0.45$ ticks for positive values of $\Delta m_0$ whereas for
negative values it is significantly larger, $-0.62$
ticks. In other words negative fluctuations have a larger knock-on effect than
positive ones, even for small fluctuations.
The total fluctuation including the initial impact, $\theta^{(m)}_{perm}(\Delta m_0)$
is shown in Figure~\ref{trueperm}.  The permanent impact has a
behavior intermediate between the zero impact assumption and the
completely permanent impact. From a linear fit of the curve for
positive and negative values of $\Delta m_0$ we obtain a 
$\theta^{(m)}_{perm}(\Delta m_0)= 0.45+0.54\Delta m_0$ (positive values) and 
$\theta^{(m)}_{perm}(\Delta m_0)=-0.62+0.48\Delta m_0$ (negative values).

In conclusion the permanent part of the fluctuation impact is roughly linear in the price (ask or mid) that is
used as a conditioning variable. The midprice permanent impact is
roughly symmetric for positive and negative fluctuations. The ask
instead shows a clearly asymmetric profile of the permanent
impact. The ask permanent impact $\Delta a_{perm}$ conditional to a
given positive ask fluctuation $\Delta a_0$ at the initial time is in
absolute value smaller than the ask permanent impact conditional to a
negative initial ask fluctuation $-\Delta a_0$. This asymmetry is a
consequence of the different causes for positive and negative ask
fluctuations, as well as of the mean reverting and positivity property
of spread.

\subsection{Transaction impact}

\begin{figure}[ptb]
  \begin{center}
\includegraphics[angle=-90,scale=0.33]{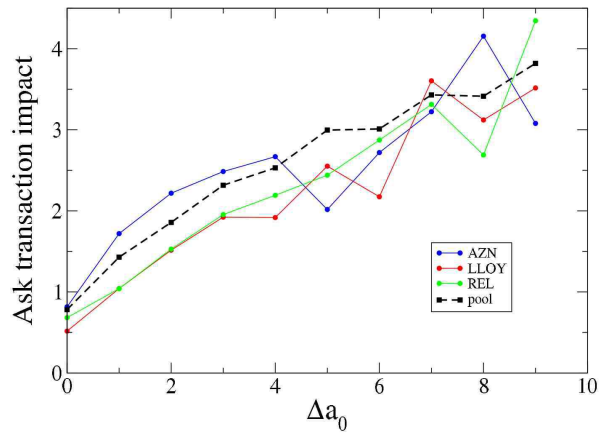} 
  \caption{Transaction permanent impact of the ask for three representative stocks and for the averaged pool across the first 55 stocks. Both axes are in tick and we show only the values corresponding to $\Delta a_0\ge 0$.}
  \label{transimp}
  \end{center}
\end{figure}

Finally, we consider the transaction impact on the ask, i.e. we
measure the permanent impact on the ask conditional to a fluctuation
of the ask and to the presence of a transaction at a previous
time. Transaction impact is important to assess the permanent effect
that a given transaction has on the price.
By using the notation of the previous section we measure the quantity
\begin{eqnarray}
%\Delta \tilde a_{perm}(\tau| \Delta a_0) = \nonumber \\
E[a(t+\tau)-a(t-1)|\Delta a_0 ~AND ~\exists ~one ~buy ~trade ~at ~t]-\nonumber \\
E[a(t+\tau)-a(t-1)]. 
\end{eqnarray}
and, as before, we average this quantity across the range of $\tau$
from $500$ to $1000$ seconds in order to have a proxy of the permanent
part.  The difference with the previously investigated fluctuation
impact is that we now condition also on the presence of at least a buy
market order at time $t$. Since buyer initiated trades can produce a
zero or a positive fluctuation of the ask, here we consider only the
values of the permanent transaction impact of the ask for non negative
values of the initial ask fluctuation, $\Delta a_0\ge 0$. The
permanent transaction impact of the ask is shown in
Fig.~\ref{transimp}.  First of all we note that also when $\Delta
a_0=0$ there is a $\sim 3/4$ tick permanent ask fluctuation. In other
words even when the buyer initiated trade at time $t$ does not change
the ask, at time $t+\tau$ the ask is on average $3/4$ ticks higher
than the ask at time $t-1$, i.e. just before the trade. This can be
due to the fact that liquidity providers raise the ask as a response
to a buy market order even if the market order itself does not create
a price change. Alternatively it is known that market order signs are
significantly correlated in time \cite{Bouchaud04,Lillo03b}. Thus even
if a specific buy market order can have vanishing immediate impact,
the market order is likely part of a wave of buy market orders that
generate a positive permanent impact.  From Fig.~\ref{transimp} we
note that for $\Delta a_0>0$ the permanent transaction impact is
roughly proportional to the initial ask fluctuation.

\section{Discussion}

In this paper, we have shown some empirical facts of limit order book
and price dynamics in double auction financial markets, in particular
the slow scale-free decay of the spread and the approximately linear
permanent market impact function.

The slow spread decay occurring after a sudden opening of the bid-ask
spread is certainly affected by the strategic placement of limit
orders inside the spread. These strategic limit order submission
procedures are performed to attain execution priority at the best ask
or bid price after temporary liquidity crises. The scale free decay of
the bid-ask spread indicates that the strategies performed might have
not a characteristic scale.

The second focus of our paper has been on the permanent price impact
induced (i) by any event altering the spread (which we call permanent
fluctuation impact) or (ii) by a trade (which we call permanent
transaction impact). Our investigations show that the permanent impact
is statistically detectable and provides relevant information for the
modeling of price formation in high frequency data both on the ask and
on the midprice. We observe that the permanent parts of the ask and
midprice fluctuation impacts and of the ask and midprice transaction
impacts are approximately linear functions of the immediate
fluctuation or transaction impact.

This proportionality could be important in the search for the origin
of fat tails in price changes. Recently \cite{Farmer04} it has been
shown that the distribution of non-zero immediate impacts $\Delta m_0$
matches the distribution of first gaps very well. This suggests that a
major determinant of the origin of large price changes is the presence
of large gaps in the limit order book when the market is in a state of
lack of liquidity. Clearly the correspondence observed in
\cite{Farmer04} holds only for individual returns (impact) and it is
not a priori obvious that one can extend it to longer time scales. The
results in Figs.~\ref{trueperm} and \ref{transimp} give support to the
idea that temporary fluctuations of the market liquidity are also
responsible for the fat tails of price changes at longer time
scales. In fact the distribution of gaps is equal to the distribution
of immediate impacts and Figs.~\ref{trueperm} and \ref{transimp} show
that permanent impact is a linear function of immediate impact.

\acknowledgments Authors acknowledge support from the research project
MIUR 449/97 ``High frequency dynamics in financial markets" and
MIUR-FIRB RBNE01CW3M ``Cellular Self-Organizing nets and chaotic
nonlinear dynamics to model and control complex system''. F.L. and
R.N.M also acknowledge support from the European Union STREP project
n. 012911 ``Human behavior through dynamics of complex social
networks: an interdisciplinary approach.''.

\end{document}